\documentstyle[11pt,newpasp,twoside]{article}
\index{summary}
\index{instructions}
\index{template}

\def\edcomment#1{\iffalse\marginpar{\raggedright\sl#1\/}\else\relax\fi}
\reversemarginpar

\begin{document}
\title{Search for the Optical Counterparts 
         of Southern Anomalous X-Ray Pulsars
         and Radio-Quiet Neutron Stars in Young Supernova Remnants}
\author{Zhongxiang Wang and Deepto Chakrabarty}
\affil{Department of Physics and Center for Space Research, Massachusetts Institute of Technology, Cambridge, MA 02139}

\begin{abstract}
We report on our search for the optical counterparts of the Southern
Hemisphere anomalous X-ray pulsar 1E1048.1-5937 and 
the radio-quiet neutron stars in supernova remnants Puppis A, 
RCW 103, and PKS 1209-52. The observations were carried out with the
new MIT/CfA MagIC camera on the Magellan-I 6.5 m telescope in Chile.
We present deep multiband optical images of the X-ray error circles for
each of these targets and discuss the resulting candidates and limits.
\end{abstract}

\section{Background}

{\bf The anomalous X-ray pulsars (AXPs)} are a group of X-ray pulsars 
whose spin periods fall in a narrow range ($\sim 6-12$ s), whose X-ray 
spectra are very soft, and which show no evidence that they accrete
from a binary companion (see Mereghetti 1999 for a recent review).
These objects may be isolated neutron stars with extremely strong
($\sim 10^{14}$ G) surface magnetic fields, or they may be accreting
from a ``fallback'' accretion disk.  Optical measurements could
potentially help discriminate between these models. 
An optical counterpart to one AXP, 4U~0142+61, has recently been
identified and shown to have peculiar optical colors (Hulleman et
al. 2000).\\
{\bf The radio-quiet neutron stars (RQNSs)} are a group of compact
X-ray sources found near the center of young supernova remnants.
Their X-ray spectra are roughly consistent with young, cooling neutron
stars, but they show no evidence for the non-thermal emission
associated with ``classical'' young pulsars like the Crab (see Brazier
\& Johnston 1999 for a review).  The X-ray spectral properties of the
RQNSs and the AXPs are similar (see, e.g., Chakrabarty et al. 2001).  
Below in Table 1, the general properties of the three RQNSs as our targets in
the southern sky are listed. 
\begin{table}
\caption{Radio-Quiet Neutron Stars}
\begin{tabular}{clcccc}
\hline
 &  & $d$ & Age & $kT_{bb}$ & \\
Source & SNR & (kpc) & ($10^3$ yr) & (keV) & Refs \\
\hline 
1E 0820--4247 & Pup A & 2.0 & 3.7 & 0.28 & 1-3  \\
1E 1614--5055 & RCW 103 & 3.3 & 1-3 & 0.56 &  4-6\\
1E 1207--5209 & PKS 1209--52 & 1.5 & 7 & 0.25 & 7-9 \\
\hline
\multicolumn{6}{l}{\tiny {\bf References.} -- (1) Petre et al. 1982. (2) Petre, Becker, \& Winkler 1996.
(3) Pavlov et al. 1999.}\\
\multicolumn{6}{l}{\tiny (4) Tuohy \& Garmire 1980. (5) Caswell et al. 1975.
(6) Gotthelf, Petre, \& Hwang 1997. (7) Helfand \& Becker 1984.}\\
\multicolumn{6}{l}{\tiny 
(8) Bignami, Caraveo, \& Mereghetti 1992. (9) Mereghetti, Bignami, \& Caraveo
1996.}\\
\end{tabular}
\end{table}
\section{Observations}
Our observations were made using the Magellan Instant Camera (MagIC)
on the Magellan-1/Walter Baade 6.5-meter telescope at Las Campanas
Observatory, Chile.  MagIC is a CCD filter photometer built by MIT and
CfA for the $f/11$ focus of the Baade telescope.  The current detector
is a 2048$\times$2048 SITe CCD with a 69~mas/pixel scale and a
142$\times$142 arcsec field of view.  We used the Sloan filter set,
which have the following central wavelengths (Fukugita et al. 1996):
$u'$=3540 \AA; $g'$=4770 \AA; $r'$=6230 \AA; $i'$=7620 \AA; and
$z'$=9130 \AA.
\begin{table}
\caption{Magellan Observations }
\begin{tabular}{lcccccc}
\hline
 &  & \multicolumn{5}{c}{Exposure Time (min)} \\
\cline{3-7}
Source & Date & $u'$ & $g'$ & $r'$ & $i'$ & $z'$\\
\hline
1E 1048.1--5937 & 2001 Mar 24 & ... & ... & 10 & 10 & 10 \\
1E 1207--5209 & 2001 Mar 24 & ... & 10 & 9 & 10 & 10 \\
              & 2001 Jun 12 & 5   & ... & ... & ... & ... \\
1E 0820--4247 & 2001 Mar 25  & ... & ... & 10 & 10 & 6 \\
1E 1614--5055 & 2001 Mar 25 & ... & ...& 10 & 10 & 10 \\
\hline
\end{tabular}
\end{table}

\section{Results}
{\bf Puppis A and PKS 1209--52}\ As shown in Figure 1, 
no optical counterparts are found in our images within the error
regions derived from {\it Chandra}/ACIS observations of these
supernova remnants.  (For PKS~1209--52, the {\em Chandra} observations
were made in CC mode, so that only a 1-D position constraint is given:
this is plotted as two parallel dashed lines in Fig. 1(b).)   The upper
limits on an optical counterpart for PKS~1209--52 are: $u' > 22.5$, $g'
> 24.4$, $r' > 24.8$, $i' > 24.4$, and $z' > 22.9$.  For Puppis~A, the
limits in $r'$ and $i'$ bands are similar to those for PKS~1209--52,
and $z' > 22.4 $. \\
\begin{figure}
\plotfiddle{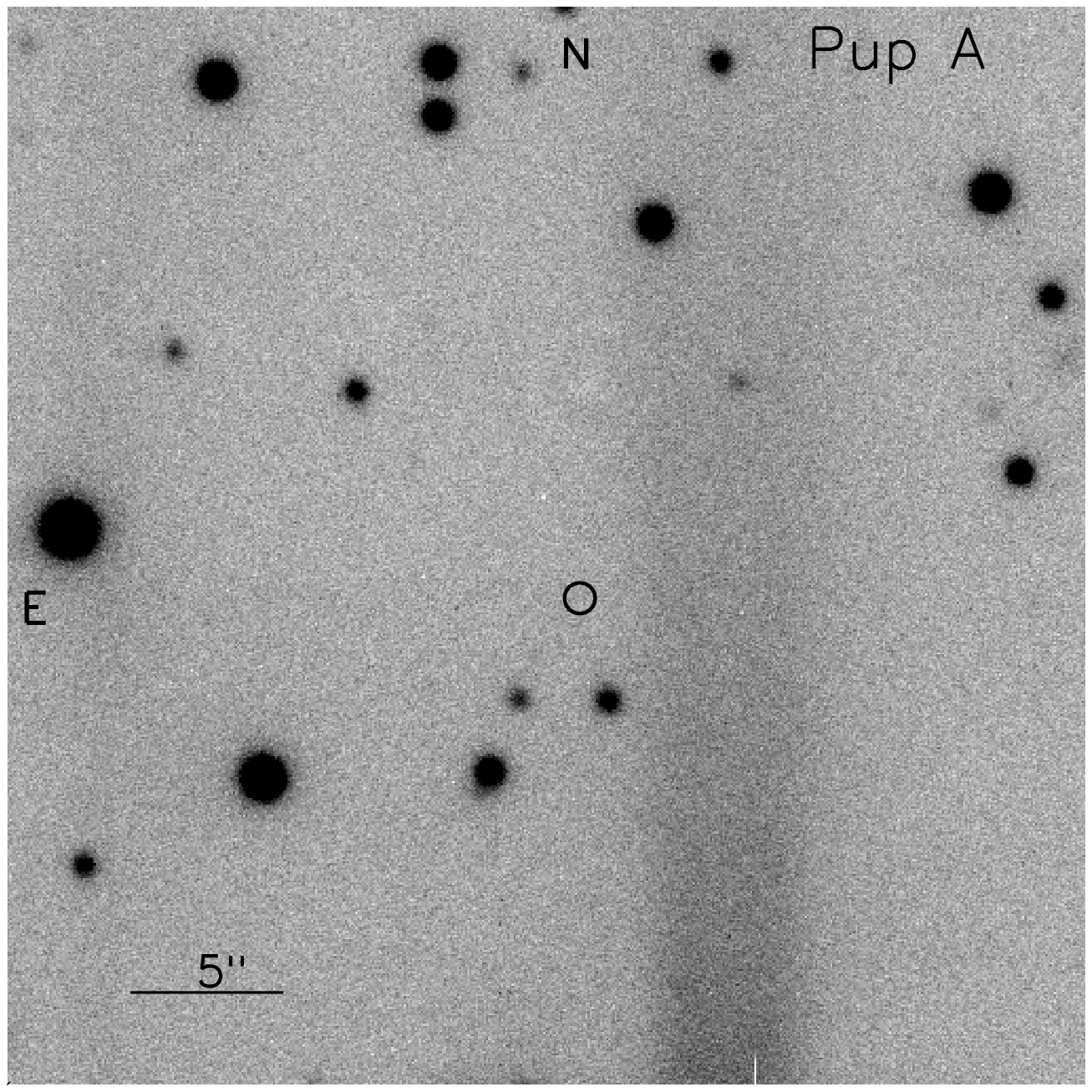}{1.0 in}{0}{041.5}{040.0}{-190.5}{-90}
\plotfiddle{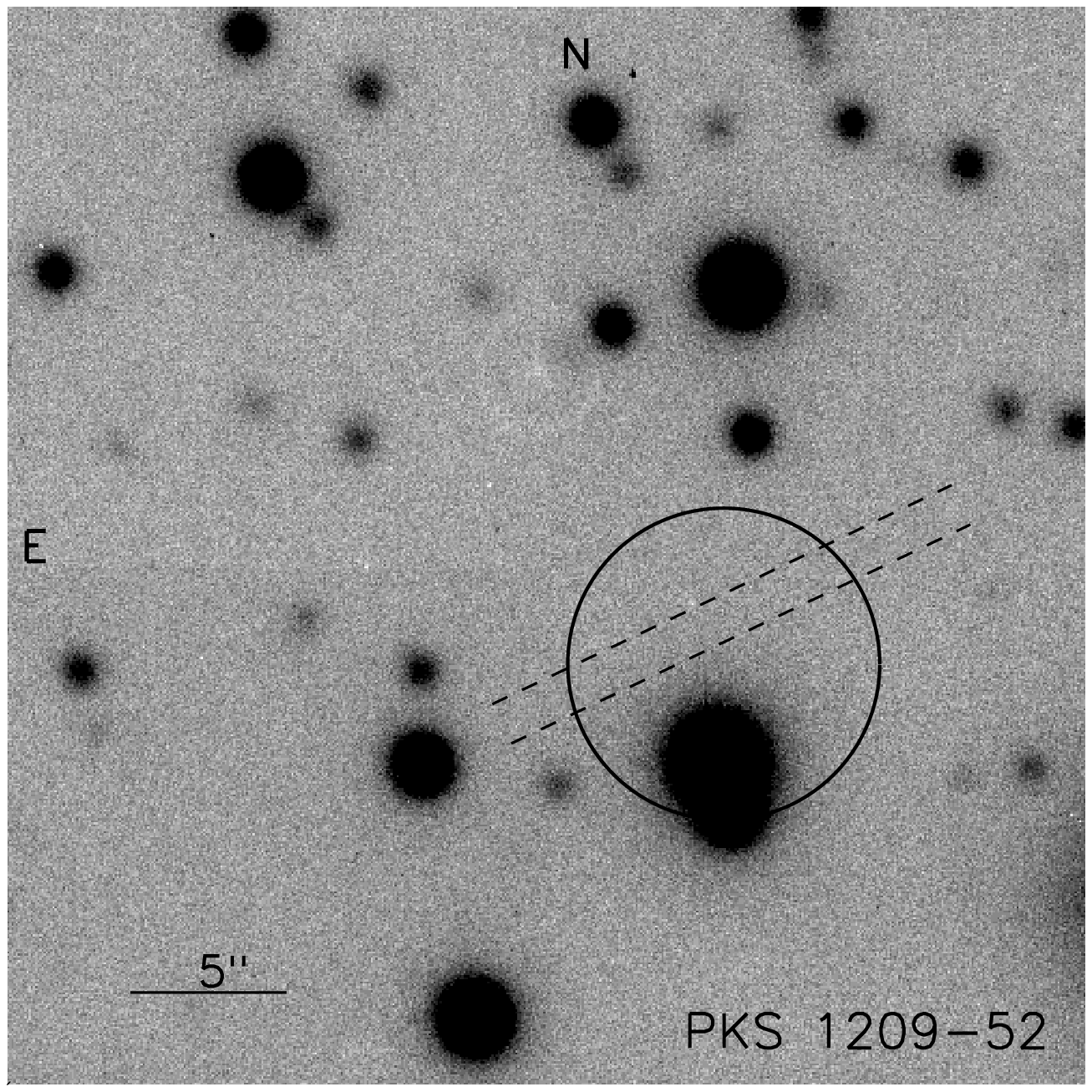}{1.0 in}{0}{041.5}{040.0}{-28.5}{-5}
\caption{{(a) Magellan MagIC image of Puppis A (1E 0820--4247) 
                  region in $i'$ band.
                 The X-ray source position, derived from the {\it Chandra}/ACIS
                 observation, is indicated as a 1$\arcsec$ diameter circle.
	(b) Magellan MagIC image of PKS 1209--52 (1E 1207--5209)
                 region in $i'$ band. A 5$\arcsec$ circle indicates the X-ray
                 source position given by {\it Einstein}/HRI observation 
                 (Seward 1990). Two parallel dotted lines with the width
                 of 1$\arcsec$.4 between them, which cross
                 the HRI error circle, are drawn to indicate the position
                 derived from {\it Chandra}/ACIS observation.}}
\end{figure}
{\bf 1E~1048.1--5937}\ As {\em Chandra} has not observed this
AXP yet, the error circle (from {\em EXOSAT}) is relatively large,
15$\arcsec$.  We are able to identify 41 stars in this region that are
bright enough to be measured in $r'$, $i'$, and $z'$ bands (see Fig.
2(a)). However, none of these stars has peculiar colors (see Fig. 2(b)).
\begin{figure}
\plotfiddle{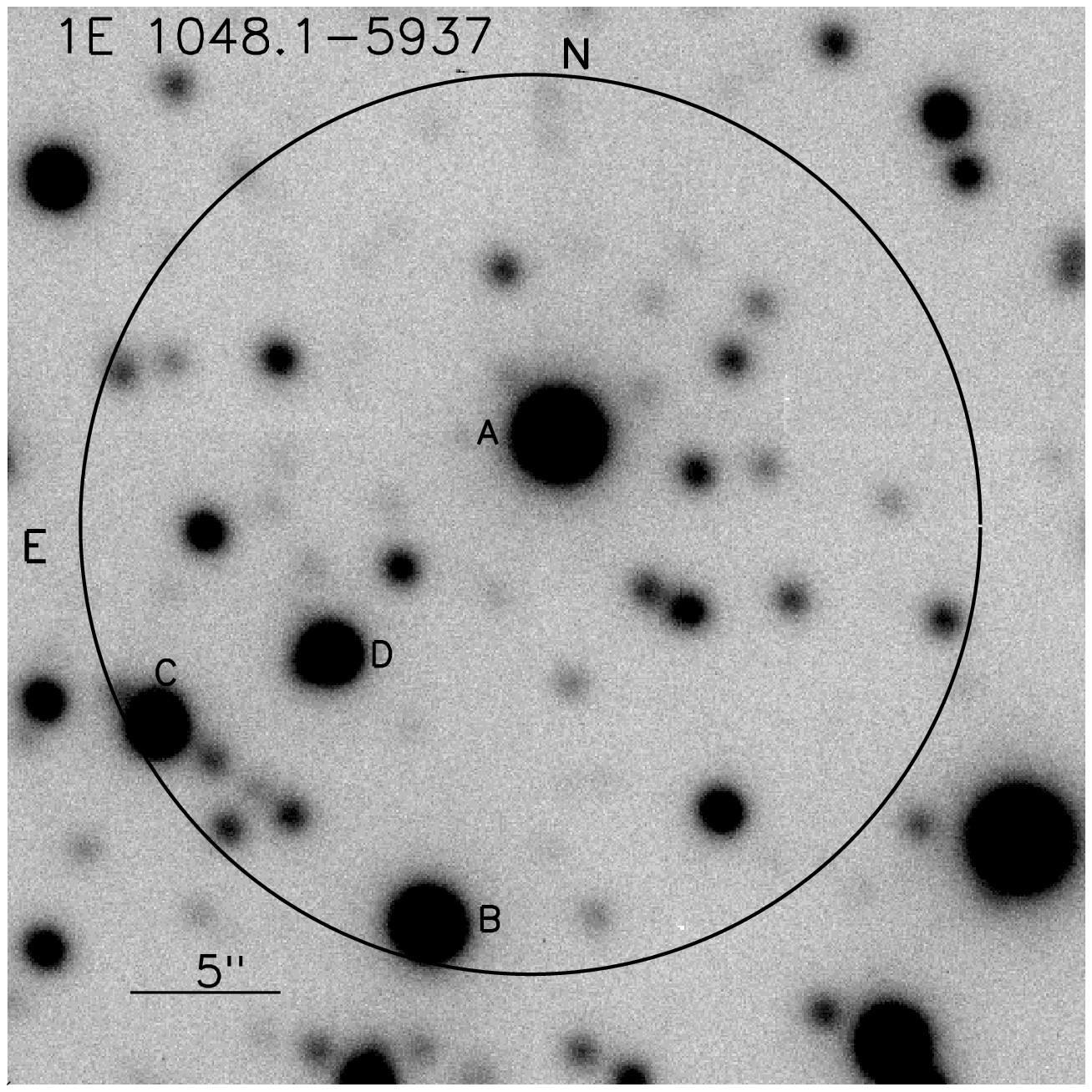}{0.9 in}{0}{041.5}{040.0}{-190.5}{-90}
\plotfiddle{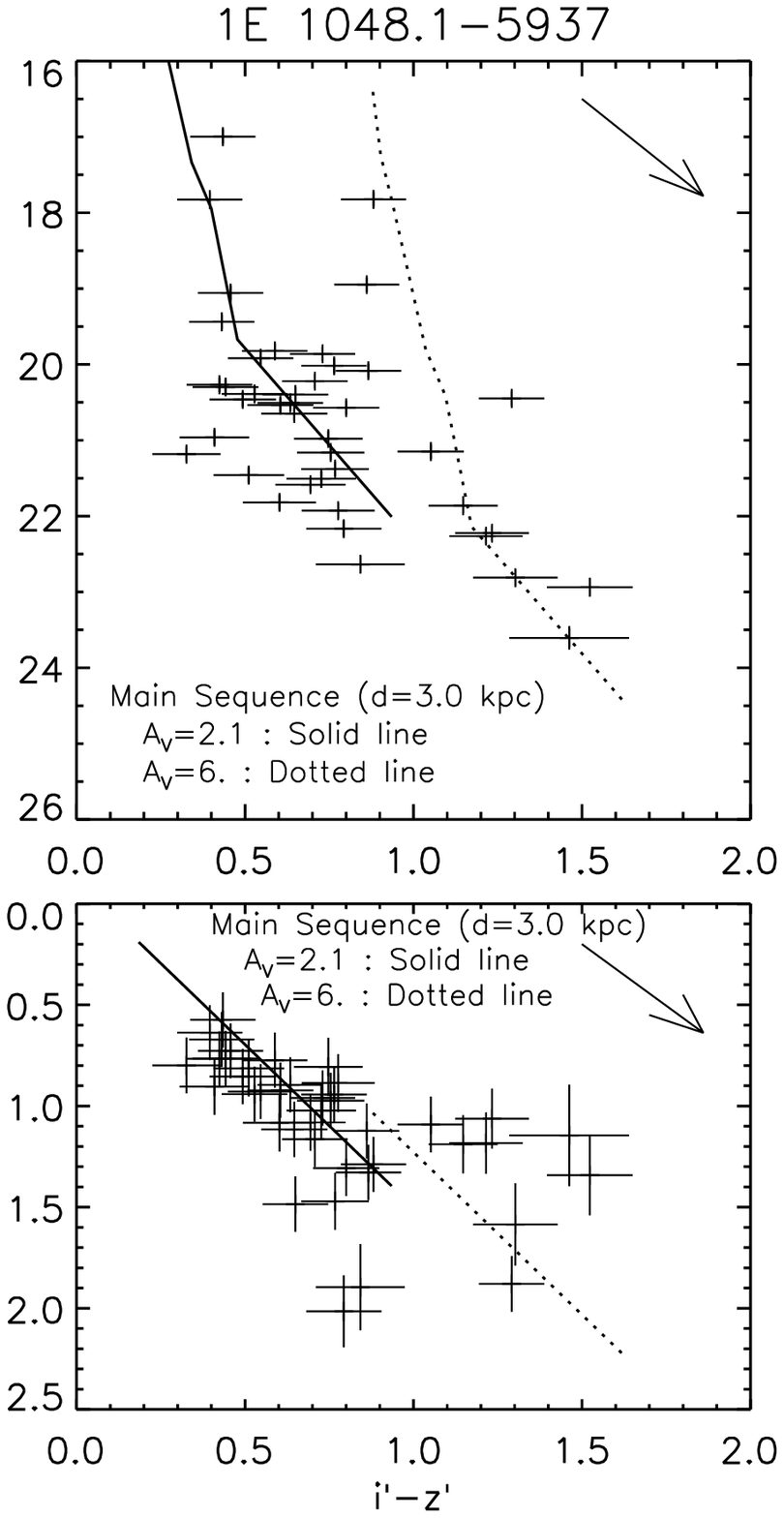}{0.9 in}{0}{031.5}{030.0}{30.5}{-9}
\caption{{(a) Magellan MagIC image of 1E 1048.1--5937 region in $i'$ band.
                A 15$\arcsec$ circle indicates the X-ray source position given
                by EXOSAT/CMA observation (Mereghetti et al. 1992). Star A, B,
                C, and D are four stars measured by Mereghetti et al. (1992;
                 see Figure 1 in their paper).
(b) Color-magnitude and color-color diagrams for objects within the error 
circle of 1E~1048.1--5937 shown in (a). The solid lines and dotted lines
indicate the main sequence with $A_{\mbox{V}} =2.1$ and $A_{\mbox{V}} = 6.$
respectively, both at a distance of 3 kpc. The arrows at the upper right corner show the reddening direction with $A_{\mbox{V}} = 2.$}}
\end{figure}
\hspace{-8mm} {\bf RCW 103}\ No optical counterpart is detected in the
0$.\arcsec54$ error circle derived from {\it Chandra}/ACIS
observations.  However, three faint stars are detected 
near the error circle ($<2\arcsec.$5 away from the X-ray position),
which we label as A, B, and C in Figure 3(a).  Their magnitudes are shown below:
\begin{quote}
\begin{tabular}{cccc}
 & $r'$ & $i'$ & $z'$ \\
A & $>$ 25.6 & 23.20 $\pm 0.15$ & 21.56 $\pm 0.04$ \\
B & 25.6$\pm $ 0.3 & 22.22$\pm $ 0.14 & 20.36$\pm $ 0.02 \\
C & 23.8$\pm $ 0.1 & 21.48$\pm $0.13 & 19.94$\pm $ 0.02 \\
\end{tabular}
\end{quote}
As shown in Figure 3(b), the colors of star B and C are normal which
are consistent with those of reddened main sequence stars. Star A may be 
a normal star with extremely high extinction.
\begin{figure}
\plotfiddle{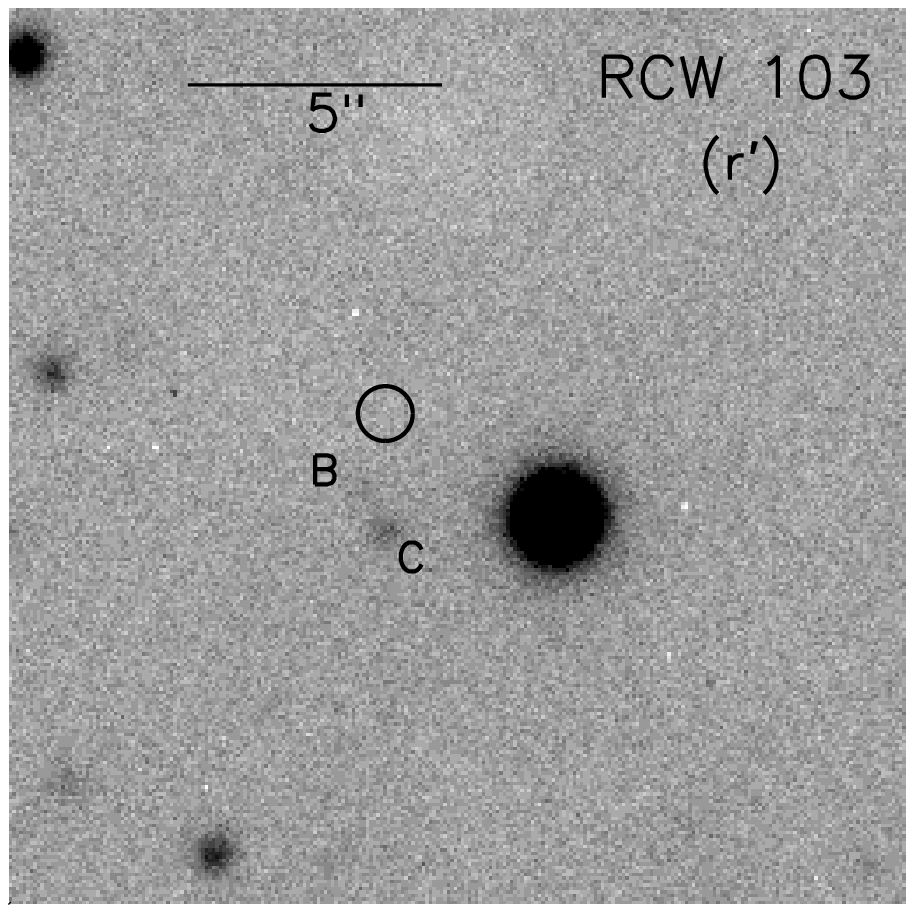}{0.4 in}{0}{031.5}{030.0}{-180.5}{-62}
\plotfiddle{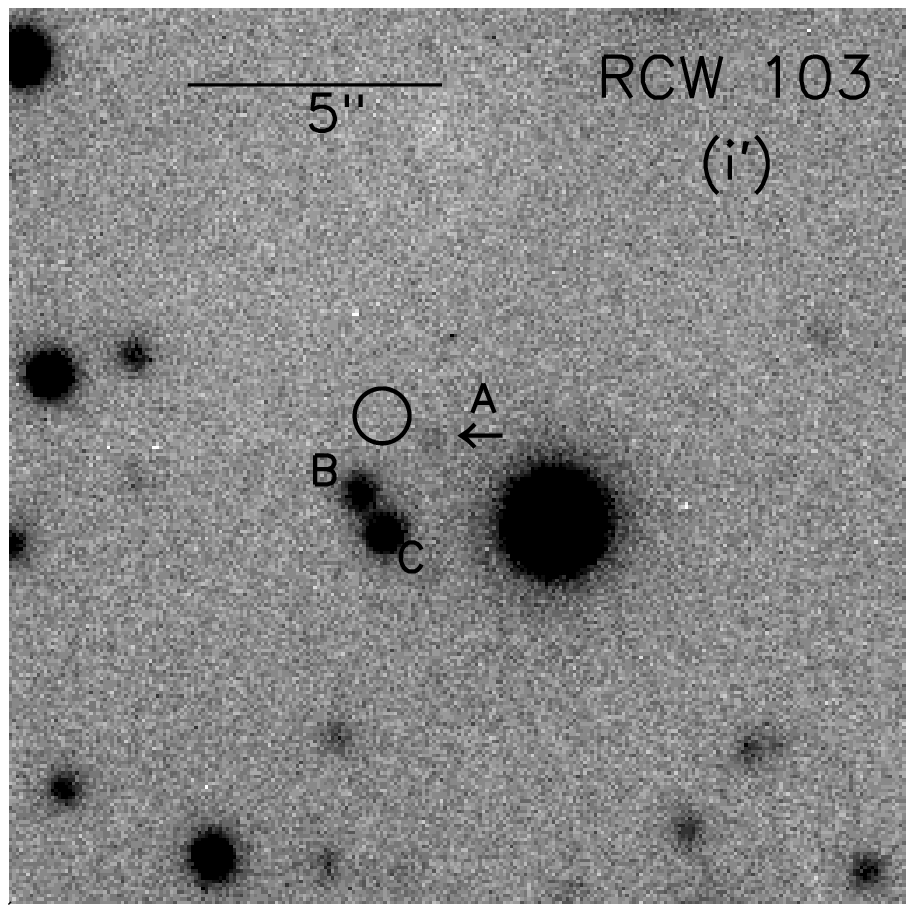}{0.4 in}{0}{031.5}{030.0}{-90.5}{-20}
\plotfiddle{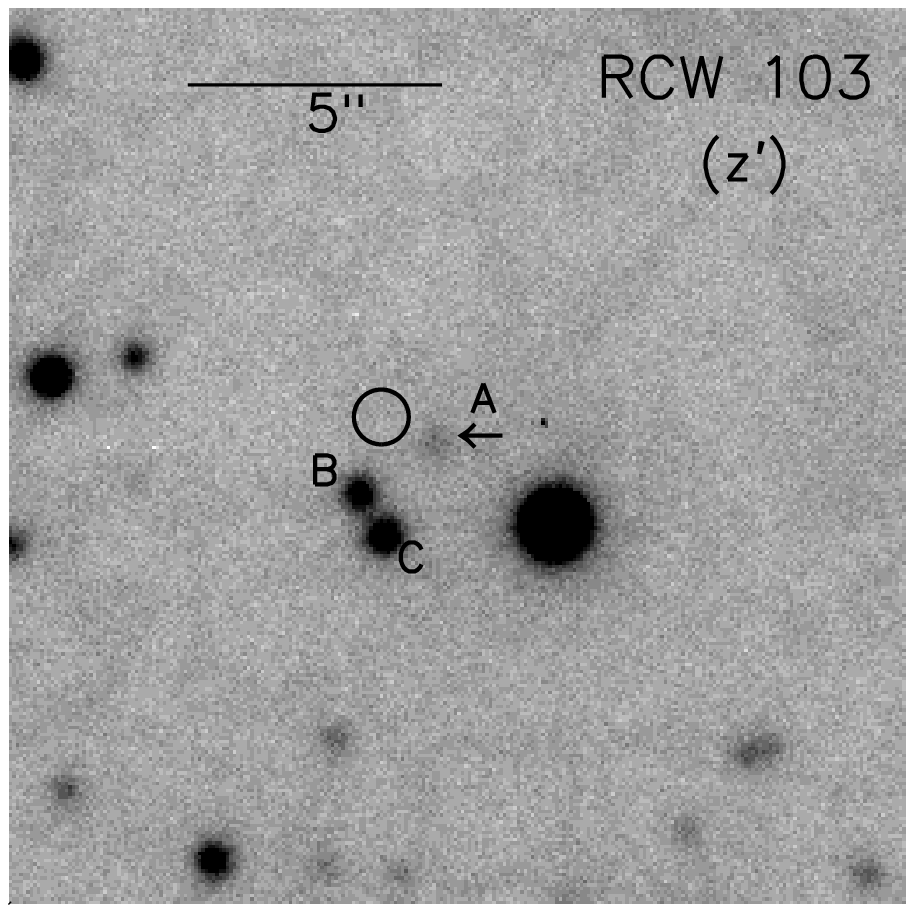}{0.4 in}{0}{031}{030.0}{-135}{-60}
\plotfiddle{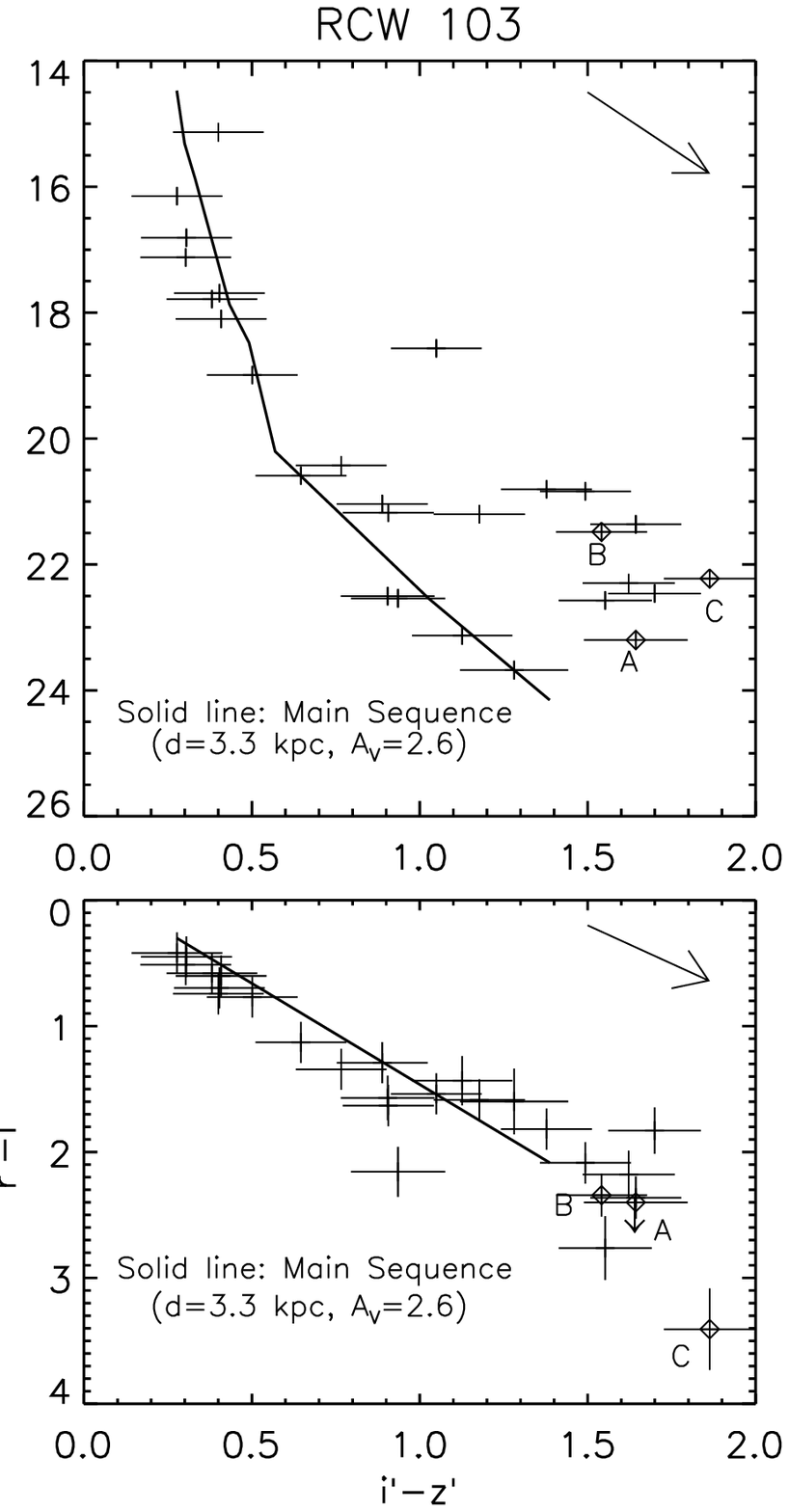}{0.7 in}{0}{031}{030.0}{50}{15}
\caption{{(a) Magellan MagIC images of the region around 
                RCW103 (1E 1614-5055) in $r'$, $i'$, and $z'$ band. The 0$\arcsec$.54
                circle indicates the X-ray position derived 
                from {\it Chardra}/ACIS
                observation (Garmire et al. 2000). 
                (b) Color-magnitude and color-color diagrams for objects near the X-ray
source position of RCW~103. The solid lines indicate the main sequence
with $A_{\mbox{V}} = 2.6$ at a distance 3.3 kpc. 
The star A, B, and C shown in Fig. 4(a) are labeled. For star A, the upper
limit of the color $r'$-$i'$ is shown. }}
\end{figure}

\footnotesize


\begin{references}
\reference Brazier, K.T.S., \& Johnston, S. 1999, MNRAS, 303, L1
\reference Bignami, G.F., Caraveo, G.A., \& Mereghetti, S. 1992, ApJ, 389, L67
\reference Chakrabarty, D. et al. 2001, ApJ, 548, 800
\reference Fukugita, M. et al. 1996, AJ, 111, 1748
\reference Garmire, G.P., Pavlov, G.G., \& Garmire, A.B. 2000, IAUC, 7350, 2
\reference Gotthelf, E.V., Petre, R., \& Hwang, U. 1997, 487, L175
\reference Helfand, D.J., \& Becker, R.H. 1984, Nature, 307, 215
\reference Hulleman, F., Kerkwijk, M.H., \& Kulkarni, S.R. 2000,
Nature, 408, 689
\reference Mereghetti, S. 1999, in The Neutron Star--Black Hole Connection,
ed. C. Kouveliotou et al., (Dordrecht: Kluwer)
\reference Mereghetti, S., Caraveo, P., \& Bignami, G.F. 1992, A\& A, 263, 172
\reference Mereghetti, S., Bignami, G.F., \& Caraveo, P.A. 1996, ApJ, 464, 842
\reference Pavlov, G. G., Zavlin, V.E., \& Tr\"{u}mper, J. 1999, ApJ, 511, L45
\reference Petre, R., Becker, C.M., \& Winkler, P.F. 1996, ApJ, 465, L43
\reference Petre, et al. 1982, ApJ, 258, 22
\reference Seward, F.D. 1990, ApJSS, 73, 781
\reference Tuohy, I., \& Garmire, G. 1980, ApJ, 239, 107
\end{references}
\end{document}